\documentclass[a4paper,11pt]{article}
\usepackage{pos}
\usepackage{floatrow}
\usepackage{xcolor}
\usepackage[normalem]{ulem}

\renewcommand{\printHeadAuthors}{J.J. G\'alvez Viruet {\it et al.}}

\begin{document}
\title{First-principle predictions of fragmentation functions via quantum computing}
\ShortTitle{Quantum computing QCD fragmentation functions }

\author*[a]{Juan J. G\'alvez Viruet, Felipe J. Llanes-Estrada }
\author[b]{Nicol\'as M. Arenaza}
\author[c]{ Mar\'{\i}a G\'omez-Rocha  }
\author[d]{T. J. Hobbs}

\affiliation[a]{Departamento de F\'{\i}sica Te\'orica\& IPARCOS, Univ. Complutense de Madrid\\
  Plaza de las Ciencias 1, Madrid 28040 Spain}

\affiliation[b]{Instituto de F\'{\i}sica Te\'orica, IFT-UAM/CSIC, 28049 Madrid, Spain}

\affiliation[c]{%
Depto. de F\'{\i}sica At\'omica, Molecular y Nuclear
and Instituto Carlos I de F\'{\i}sica Te\'orica y Computacional,
Universidad de Granada, 18071 Granada, Spain
}%

\affiliation[d]{High Energy Physics Division, Argonne National Laboratory, Lemont, IL 60439, USA}

\emailAdd{fllanes@fis.ucm.es}
\emailAdd{juagalve@ucm.es}

\abstract{We report on an algorithm to compute fragmentation functions from the first principles Quantum Chromodynamics (QCD) Hamiltonian quantized in Light-Front Gauge, opening a path for digital quantum computers to calculate 
these longitudinal jet-structure observables. 
Simulating the behaviour of such computers on a classical cluster (which is memory-limited to about 30 qubits, given the expansive Hilbert spaces of actual quantum computers), we run a demonstration 
of a heavy-quark leading parton fragmenting into quarkonium, which we benchmark against NRQCD computations. Future quantum computers, perhaps concurrently running with HL-LHC, would have ample opportunity to extract arbitrary parton-hadron combinations.
}

\FullConference{14th Edition of the Large Hadron Collider Physics (LHCP2026)\\
18-22 May 2026\\
Paris, France\\}


\maketitle

\section{Introduction: when will quantum computers become useful for theory?}

Quantum computing is making remarkable progress towards being a useful tool in hadron physics~\cite{Galvez-Viruet:2025ket} but it is not quite yet at the point where it can fill the gaps in our understanding of the strong force that lattice gauge theory currently struggles with. There is, however, promise.

\begin{figure}[h]
\floatbox[{\capbeside\thisfloatsetup{capbesideposition={left,top},capbesidewidth=4cm}}]{figure}[\FBwidth]
{\caption{Progress in quantum computing (symbols on the lower left, from practical published calculations) towards meeting competitive computations for fragmentation function (beyond the orange lines). Also recorded, at the top right, is a resource estimate for breaching standard RSA cryptography. The number of usable qubits has already exceeded the dashed line beyond which classical RAM memory cannot represent the $2^n$-dimensional Hilbert space vectors.}
\label{fig:resources}}
{\includegraphics[width=0.7\textwidth]{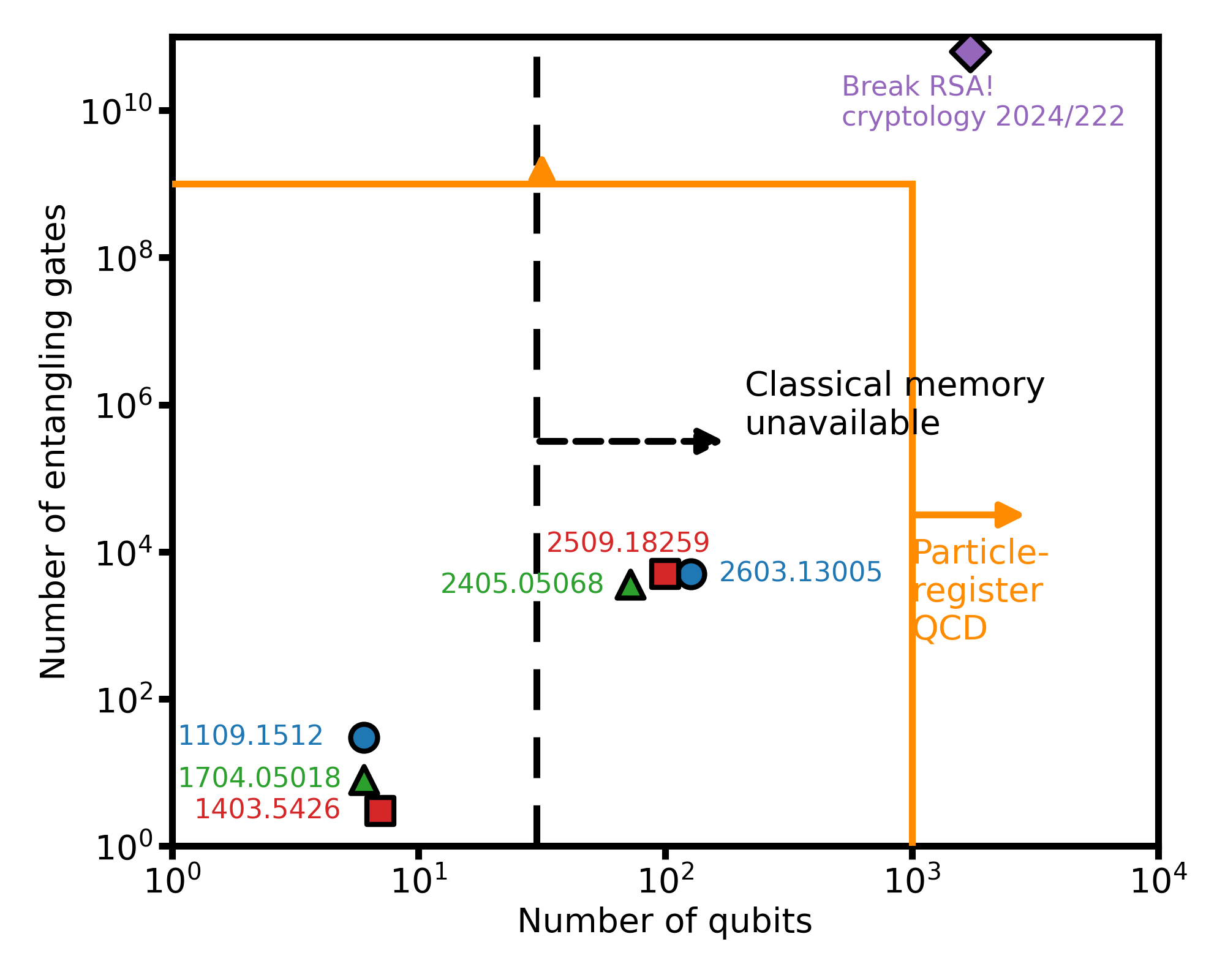}}
\end{figure}

In Fig.~\ref{fig:resources} we show our estimate of the resources (number of qubits, number of entangling gates needed) to make credible calculations of fragmentation functions.
Good progress has been made on the number of qubits (see data points with various preprint numbers), which is reaching in practical calculations the $100$ count (chips with $1000$ qubits are at hand but computations with such large numbers are not yet productive due to modest connectivity between qubits, decoherence and noise). The number of entangling gates which have realistically been computed before noise takes over, however, is of order $10^4$ as of 2026, when $10^9$ would be desirable, so there is still a need for important improvements in the reliability of quantum computations.

While the plot makes clear that we are still far from being able to deploy serious-scale hadron calculations on quantum computers, it is well possible that the orange lines will be exceeded in both dimensions (number of qubits and number of entangling 2-qubit gates) while the HL-LHC  as well as the Electron-Ion Collider at Brookhaven are operating in the 2030s and beyond.

In that case, a program of prediction of hadron-structure distributions and pseudodistributions newly computed from first principles and comparison with experimental data could be envisioned, even in sectors such as fragmentation functions still resisting conventional lattice efforts.

In spite of extensive discussion on what "quantum advantage" is and when has it been/will it be achieved, we simply surmise that current quantum chips already span massive Hilbert spaces $\alpha_1 |1\rangle + \cdots 2^{\rm n\ qubits}|2^n \rangle$ which cannot be stored on any classical RAM memory. Therefore, in many-body problems, useful work is already possible if the number of operations is modest. This will not be the case for Chromodynamics, whose Hamiltonian, to which we now turn, is very complicated and requires large numbers of gates.

\newpage
\section{Particle-register encoding for QCD in the light-front}

\begin{figure}[b]
    \centering
\includegraphics[width=0.9\linewidth]{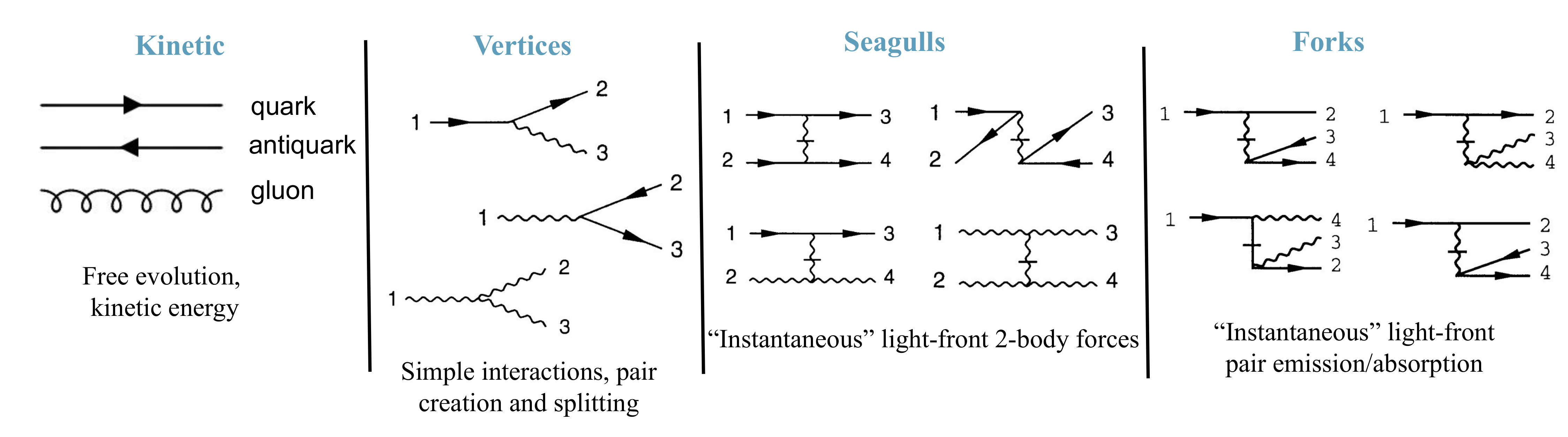}
\includegraphics[width=0.9\linewidth]{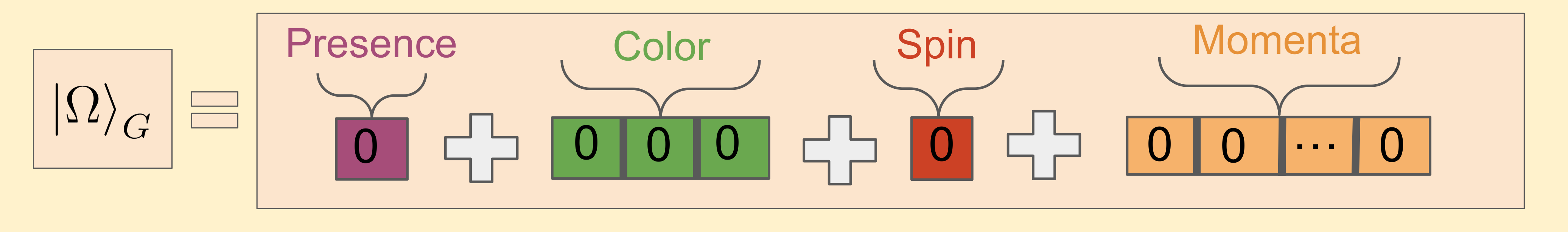}
    \caption{{\it Upper plot}: typical terms of the Light-Front QCD Hamiltonian~\cite{Brodsky:1997de} (fully developed algebraic expressions in terms of particle creation/destruction operators are involved~\cite{Galvez-Viruet:2025rmy}). {\it Lower plot}: an empty gluon-register (setting the "presence" qubit to 1 fills it with one gluon of the quantum numbers specified by the following qubits). These registers are entangled by the (particle-number changing) Hamiltonian terms.
    }
    \label{fig:encoding}
\end{figure}

Computations with Quantum Chromodynamics in Light-Front gauge $A^+=0$ entail passing from the well-known QCD Lagrangian, via the Legendre transform, to the Hamiltonian
\begin{equation}
P^- = \sum \Pi_i \frac{\partial}{\partial x^+} A_i -\mathcal{L}_{\rm QCD}\ .
\end{equation}
A few typical terms of this Hamiltonian are given in the top plot of Fig.~\ref{fig:encoding}. As a second-quantized system, the interactions can change the particle number. Our philosophy~\cite{Galvez-Viruet:2024hry,Galvez-Viruet:2026uaw} consists of implementing the particle creation-destruction operators $a$, $a^\dagger$
as computer methods (programmed in Python and Qiskit) acting on memory registers. For example, if $a^\dagger$ acts on a gluon-register which is empty (lower plot of Fig.~\ref{fig:encoding}), it flips the Absence/Presence qubit to 1 and writes the quantum numbers of the new gluon on the following qubits.

At any given time we want the memory of the quantum computer (which represents the wavefunction of a system entangling different quantum numbers and different numbers of particles) to be (anti)symmetrized upon exchange of identical (fermions) bosons. This is achieved by means of appropriate step (anti)symmetrizers which are applied upon creating any particle to guarantee its correct symmetry properties {\it vis à vis} the already occupied particle registers of the same species. 

The programmed particle creation/destruction operators satisfy canonical commutation rules, for gluonic modes \begin{equation}[a_i,a_j^\dagger]= \delta_{ij}+{\rm boundary}\ ,
\end{equation} 
the boundary term activates when the memory is full (so $a^\dagger$ cannot step up the particle number as there are no more qubits to register the new quantum).

This encoding is a generic tool to approximate quantum field theories on a quantum computer, distinct from lattice gauge theory and Jordan-Wigner approaches, {\it e.g.} ~\cite{Li:2024nod}~within the Nambu-Jona-Lasinio model
and we here apply it to the computation of a fragmentation function.

\newpage
\section{A demonstration of the algorithm}

We now turn to the main result of this investigation, a demonstration of feasibility to extract fragmentation functions from a quantum computer. Our largest classical-cluster simulation entails 29 qubits for color $SU(3)$, and does not permit fully exploring cutoff dependences, extrapolations in the Trotter step or including transverse momentum, while also having a limited number of particles (two quarks, one antiquark and one gluon in the jet) and longitudinal-momentum grid. To eliminate these restrictions we should be able to run the computation in machines with order $10^3$ qubits or more, obviously unrepresentable in current classical simulators. 

Fragmentation functions were introduced early on upon noticing the existence of hadron jets and their parton-level meaning~\cite{Gronau:1973gc,Feynman:1973xc}.
They encode the probability to extract a hadron $h$ independently of any other particles in the event, $X_{\text{out}}$, out of the jet with initial parton $j$. The longitudinal momentum fraction is then  $z=p_h^+/p_j^+$. 
At a fixed scale~\cite{Collins:2023cuo}, the fragmentation function is

\begin{eqnarray}
    D_{j}^h(z) 
    \equiv \frac{\rm Tr_{s}  \rm Tr_{c} } {3\,N_{c,j}}\sum_X 
    \langle j,p_j | h_s,X_{\text{out}}\rangle 
    \langle h_s,X_{\text{out}} | j,p_j\rangle 
     \delta(z - p_h^+/p_j^+) \ ,\label{FuncionFragmentacion}
\end{eqnarray}
with $N_{c,j} $ the number of, and the trace taken over the parton colors, respectively.

We choose to simulate a charm-initiated jet and to extract a $J/\psi$ with approximate wavefunction (an ansatz as diagonalizing $H_{QCD}$ would be a separate effort)
\begin{eqnarray} \label{Jpsiwf}
|J/\Psi\rangle =  \sum \delta_{ c_qc_{\bar{q}} }  \frac{\chi_0(x) \,\vec{\sigma}_{ij}}{\sqrt{x \,(z-x)}}
\left|x\ i\,c_q ,(z-x)\ j\,c_{\bar{q}}\right\rangle \ , \ \ 
\end{eqnarray}
(the quark and antiquark momentum fractions are here $x<z$ and $z-x$), 
The longitudinal wavefunction is a simple ansatz~\cite{Li:2021cwv}, with $C$ normalizing to unity over $x\in [0,z]$ and physical parameters
$\alpha = 4 m_q^2/\kappa$, $\beta = 4 m_{\bar{q}}^2/\kappa$, with $m_q = m_{\bar{q}} = 1.27$ GeV and $\kappa = 1.34$ GeV:  
\begin{equation} \label{Cisneroschi}
\chi_0(x) = \frac{1}{\sqrt{C}} \  x^{\beta/2} (z - x)^{\alpha/2}\ .
\end{equation}

\begin{figure}[t]
\floatbox[{\capbeside\thisfloatsetup{capbesideposition={right,top},capbesidewidth=0.28\textwidth}}]{figure}[\FBwidth]
{\caption{Our evaluation of the $c\to J/\psi$ fragmentation function (data points) is compared to the extant NRQCD computations (lines). T: the totality of the Hamiltonian was included. V: partial calculation where only the potential terms of ``V'' type were included. Unlike in our earlier publication~\cite{Galvez-Viruet:2025rmy}, no very small Hamiltonian matrix elements were cut off to speed up.
\label{fig:fragmentation}
}}
{\includegraphics[width=0.7\textwidth]{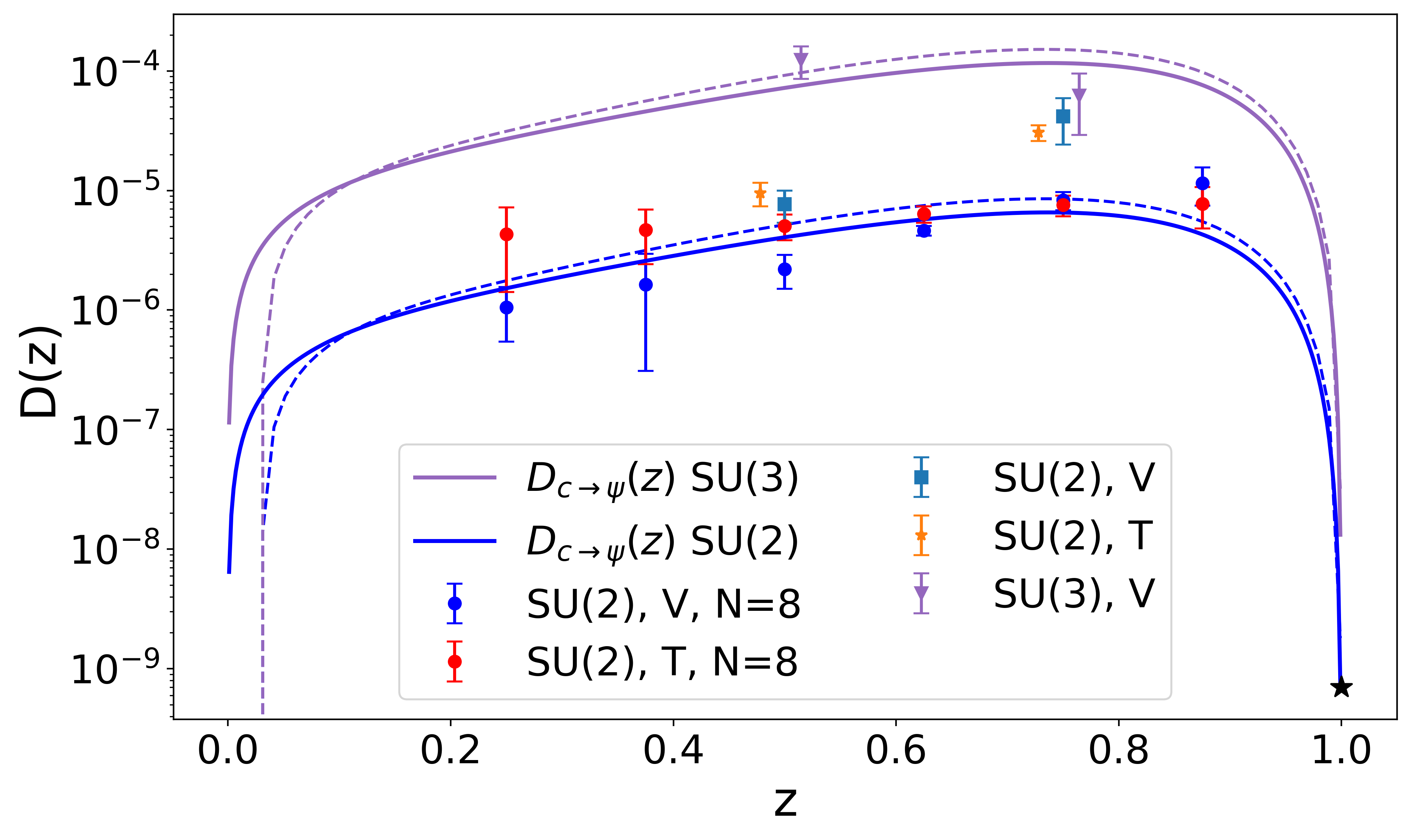}}
\end{figure}

This choice of computing a heavy-to-heavy fragmentation function is motivated by the possibility of directly comparing with reliable 
NRQCD computations~\cite{Braaten:1993mp} at a scale $\mu=3m_c$ (extended to NLO~\cite{Zheng:2019dfk}),
as done in Fig~(\ref{fig:fragmentation}) for both 2 and 3 colours.

There are several systematic uncertainties which were recently assessed in our primary publication: 
the truncation of the Fock space to the sectors of 1, 2, 3 and four particles; the discretization error introduced by the Trotter steps; the uncertainty due to truncating off very small Hamiltonian terms such that $V< \varepsilon V_{\rm max}$ {which is here lifted, as we do not discard any matrix elements, no matter how small; the uncertainty due to the scale choice; the quality of the wavefunction ansatz; and the choice of time at which the measurement is performed, decided by a plateau on the entropy of the total state}. We strove to estimate all these and give corresponding raiser uncertainty bars in Fig~(\ref{fig:fragmentation}). Within those large uncertainties, qualitative agreement with the known NRQCD computations is reasonable.

 \newpage
\section{Discussion}

 By our estimate, interesting computations for hadron structure complementing Lattice Gauge Theory would require some $10^9$ entangling gates successfully operated over order $10^3$ qubits (see Fig.~\ref{fig:resources}). This would allow to explore larger number of particles (we are currently limited to three or four partons given the large Fock spaces required
 by their many degrees of freedom), and also include transverse momentum, currently left out.
 One may ask why these requirements look smaller than those with better known encodings and lattice formulations. For example, a recent estimate of the cost in a lattice of 10 sites per dimension~\cite{Ciavarella:2025bsg},
 to order $1/N_c$ in a large-$N_c$ expansion, costs $1.2\times 10^4$ qubits and $2.8\times 10^{10}$ T-gates (45 degree rotations) per Trotter step. 

 Indeed, our encoding~\cite{Galvez-Viruet:2024hry} is economic if a few, say $n$, partons are active, with $N_p$ modes: the necessary number of qubits needed is only $n\log_2 N_p$, while widely used Jordan-Wigner encodings require $O(N_p)$ qubits although most of those modes are, in a system far from saturation, empty and unused.  Our approach would start being more costly when the number of partons became large, say at very low Bjorken $x$, losing its advantage.

We additionally note that the register wave function enfolds the reduced density matrix of the final-state hadron, alongside corresponding access to quantum concurrence, discord, and magic in the jet; these provide complementary probes of hadron structure~\cite{Bloss:2026yrf} otherwise difficult to access via Euclidean lattice approaches.

In summary, we have developed a way of computing fragmentation functions from first principles in light-front QCD with a particle-register encoding which is a natural language for few-body problems with variable particle number, and are looking forward to quantum computers becoming powerful enough to execute it in full.  For the time being, we have a small-scale demonstration running the equivalent of 29 qubits on a classical cluster.

\newpage
\section*{Acknowledgments}
FJLE thanks the organizers of the LHC Physics Conference 2026 at the Jussieu campus for their efforts;
JJGV the hospitality of the Theoretical High Energy Physics Group at Argonne National Laboratory to conduct parts of this investigation.
Supported by grants 
PID2022-137003NB-I00, 2025-170350NB-I00, 
PID2023-147072NB-I00; 
FPU21/04180 
and FPU24/00948 
of the Spanish MCIN/AEI/10.13039/501100011033/ and Ministry of Universities;
``Ayudas de Máster IPARCOS-UCM/2024'';  
and the  U.S. Department of Energy under contract DE-AC02-06CH11357.
The numerical codes were primarily run at the Carlos I institute's PROTEUS supercomputer in Granada.




\begin{thebibliography}{99}

\bibitem{Galvez-Viruet:2025ket}
J.~J.~G{\'a}lvez-Viruet, F.~J.~Llanes-Estrada and M.~G{\'o}mez-Rocha,
Int. J. Mod. Phys. A \textbf{41} (2026), 2630003
doi:10.1142/S0217751X26300036.

\bibitem{Galvez-Viruet:2024hry}
J.~J.~G{\'a}lvez-Viruet and F.~J.~Llanes-Estrada,
Phys. Rev. D \textbf{110} (2024), 116018
doi:10.1103/PhysRevD.110.116018


\bibitem{Galvez-Viruet:2026uaw}
J.~J.~G{\'a}lvez-Viruet and N.~Mart{\'\i}nez de Arenaza,
Acta Phys. Polon. Supp. \textbf{19} (2026), 4-A8
doi:10.5506/APhysPolBSupp.19.4-A8


\bibitem{Brodsky:1997de}
S.~J.~Brodsky, H.~C.~Pauli and S.~S.~Pinsky,
Phys. Rept. \textbf{301} (1998), 299-486
doi:10.1016/S0370-1573(97)00089-6


\bibitem{Galvez-Viruet:2025rmy}
J.~J.~G{\'a}lvez-Viruet, F.~J.~Llanes-Estrada, N.~M. Arenaza, M.~G{\'o}mez-Rocha and T.~J.~Hobbs,
[arXiv:2510.18869 [hep-ph]].  Accepted in Physical Review D, in production.




\bibitem{Gronau:1973gc}
M.~Gronau, F.~Ravndal and Y.~Zarmi,
Nucl. Phys. B \textbf{51}, 611-627 (1973)
doi:10.1016/0550-3213(73)90535-X


\bibitem{Feynman:1973xc}
R.~P.~Feynman,
``Photon-hadron interactions,'' 
 W.A. Benjamin, Reading, Mass. USA (1972).



\bibitem{Collins:2023cuo}
J.~Collins and T.~C.~Rogers,
Phys. Rev. D \textbf{109}, 016006 (2024)
doi:10.1103/PhysRevD.109.016006


\bibitem{Li:2024nod}
T.~Li, H.~Xing and D.~B.~Zhang,
eprint [arXiv:2406.05683 [hep-ph]].


\bibitem{Li:2021cwv}
Meijian~Li {\it et al.},
Eur. Phys. J. C \textbf{82},  1045 (2022)
doi:10.1140/epjc/s10052-022-10988-5


\bibitem{Braaten:1993mp}
E.~Braaten, K.~m.~Cheung and T.~C.~Yuan,
Phys. Rev. D \textbf{48}, 4230-4235 (1993)
doi:10.1103/PhysRevD.48.4230

\bibitem{Zheng:2019dfk}
X.~C.~Zheng, C.~H.~Chang and X.~G.~Wu,
Phys. Rev. D \textbf{100}, 014005 (2019)
doi:10.1103/PhysRevD.100.014005

\bibitem{Ciavarella:2025bsg}
A.~N.~Ciavarella, I.~M.~Burbano and C.~W.~Bauer,
Phys. Rev. D \textbf{112} (2025), 054514
doi:10.1103/ylqb-phv5


\bibitem{Bloss:2026yrf}
H.~Bloss, T.~J.~Hobbs and N.~McGinnis,
e-print ANL-205269,
[arXiv:2607.28724 [hep-ph]],     






\end{thebibliography}
\end{document}